\documentclass[aps,floatfix,nofootinbib,showpacs]{revtex4-1}

\usepackage{graphicx,color}
\usepackage{epsfig}

\usepackage{graphicx}
\usepackage{dcolumn}
\usepackage{bm}

\begin{document}



\title{An Iterative method for Analysis of Hadron Ratios and Spectra in Relativistic  Heavy-ion Collisions}

\author{Suk Choi$^1$ , Kang Seog Lee$^2$\footnote{Correspond to kslee@chonnam.ac.kr}}
 
\affiliation{$^1$ Rare Isotope Science Project, Institute for Basic Science, Daejeon, South Korea \\
$^2$ Department of Physics, Chonnam National University, Gwangju 500-757, Korea}

\date{May 15, 2014, revised on Jan. 25, 2016}

\begin{abstract}

A new iteration method is proposed for  analyzing both the multiplicities and the transverse momentum spectra measured within a small rapidity interval and the low momentum limit without assuming the invariance of the rapidity distribution under the Lorentz-boost and applied to the hadron data by the ALICE collaboration in Pb+Pb collisions at  $\sqrt{s_{NN}} = 2.76$ TeV. In order to correctly consider the resonance contribution only to the small rapidity interval measured,  ratios involving only those  hadrons whose transverse momentum spectrum is available are considered. 
In spite of the small number of ratios considered, the quality of  fitting  both of the ratios and the transverse momentum spectra is excellent. Also the calculated  ratios involving strange baryons with the fitted parameters agree with data surprisingly well.

\end{abstract}

\pacs{24.10.Pa,25.75.-q }

\maketitle

%

\section{INTRODUCTION}

In the relativistic heavy-ion collisions hadrons produced are expected to carry information on the dynamic evolution of the quark-gluon plasma and thus quantities  such as the numbers, the transverse momentum spectra,  rapidity distributions, and  the angular distributions, etc are measured in great detail.  While hydrodynamic models developed so far are very successful in explaining many of them in terms of the initial conditions obtained just after the collisions, the statistical model and the blast-wave model are also successful in fitting the multiplicities and the slopes of momentum spectra, respectively, giving insights into the  hot system of hadrons at freeze-out.

Statistical models\cite{braun1,andro,becat2,manninen,rafelski,pbm,becat,cleymans}  are known to fit various hadron multiplicities measured in relativistic heavy-ion collisions  very well with a few parameters such as the temperature, the baryon and strangeness chemical potentials, and  the success of the fitting is interpreted as the evidence of the chemical equilibration of the quark-gluon plasma formed during the collision. Usually the hadron multiplicities only within a narrow rapidity interval near $y=0$ are measured and analyzed assuming  the invariance of the rapidity distribution, $dN/dy$, under the Lorentz boost. Applying the same method to the particles in different rapidity bins one gets different temperatures for different $y$ bins.

Also the so-called blast-wave model\cite{siemens,lee,sollfrank,dobler} is very successful in fitting the transverse momentum spectra of various hadrons in terms of the temperature and the transverse expansion velocity. One draw-back of the blast-wave model is that it fits only the slopes of the transverse momentum spectra, but not the absolute magnitude of the spectra. Thus one has to use  the different normalization factor for different particle species, which basically means that the blast-wave model cannot explain the multiplicities of each hadrons. In spite  of the draw-back,  the fitting of the slopes of different hadron species with a common temperature and the transverse expansion velocity is regarded as the evidence of the hydrodynamic evolution of the matter created in the relativistic heavy-ion collisions.
 
The resulting temperature from the chemical analysis using statistical model, $T_{ch}$, is  different from that obtained in the thermal analysis of the momentum spectra using the blast-wave model, $T_{th}$. In recent hydrodynamic calculations, this difference in the freeze-out temperatures is simulated by introducing the processes  called as hadronic after-burner using UrQMD\cite{song,jeon}  or hadronic cascade\cite{hirano1}. At a certain temperature called as the switching temperature, $T_{sw}$ hadrons are generated by Monte Carlo simulation from each fluid cells.  The hadrons thus generated  undergo secondary collisions and decays via UrQMD or hadronic cascade. 
The hadronic after-burner, in some sense, takes care of the difference of $T_{ch}$ and $T_{th}$ by introducing another parameter $T_{sw}$. 

The two different freeze-outs can be implemented in  a single blast-wave model as is done in Ref.\cite{choi}. The cylindrically expanding fireball chemically freezes out at higher temperature, $T_{ch}$, where in-elastic collisions  among hadrons become less frequent to keep up the species changing reactions. The system further cools down until  $T_{th}$, where the expansion time scale becomes larger than  the characteristic time for  elastic collisions.
This model fits both the hadron ratios and the transverse momentum spectra of measured hadrons very nicely, when applied to the RHIC data.

Only the overall multiplication factor which is common to all the particle species has to be varied when the momentum spectra are fitted and thus the difference of this model with the ordinary blast-wave model is that the absolute magnitudes of the momentum spectra of various hadrons  are not adjusted to fit spectra of particle species, but they are determined already in the chemical analysis. Also this calculation is different from the  usual statistical model because it assumes the uniform temperature all over the fireball without the invariance under the Lorentz-boost, which is essential in the midrapidity region in usual statistical model.

In this article it is pointed out that there is inconsistency in extracting the number of hadrons from the transverse momentum spectra and suggest an iterative method between the chemical and thermal analysis to remove the inconsistency. The number of hadrons of a certain species, {\it e.g.} pions, is obtained from the transverse momentum spectra measured in the range $p_T> p_{T,0}$  by extrapolating to zero $p_T$  using the blast-wave model including the resonance contribution. Near zero $p_T$ the nmber of resonance decayed hadrons are very large and thus accurate calculation of the number in the small $p_T$ region  is critical in the chemical analysis. Dynamically the number of  hadrons is fixed at $T_{ch}$ including the numbers of resonances and the transverse momentum spectra is determined at $T_{th}$ includng the resonance decay. Thus in the blast-wave model the momentum spectra of hadrons decayed from resonances at $T_{th}$ should be calculated using the numbers of resonances which is fixed at $T_{ch}$. However, in the blast-wave model, the numbers of resonances are approximated by the values at $T_{th}$, which results in a systematic error in the chemical analysis. 

In this article an iterative method of analyzing  both the hadron numbers and their transverse momentum spectra  is presented within a blast-wave model\cite{choi} which assumes that the chemical freeze-out occurs at higher temperature, $T_{ch}$, and the thermal freeze-out occurs later at lower temperature, $T_{th}$. 
The method has been  applied to the hadron data\cite{alice1,alice2} by the ALICE collaboration in Pb+Pb at $\sqrt{s_{NN}} = 2.76 $ TeV and show that the result of fitting is very good without any need for strangeness fugacity or any deviation for ratios involving  protons. The fit is done only for the hadrons whose transverse momentum spectra are available, {\it i. e.} K$^{\pm}$/$\pi^{\pm}$, p/$\pi^{\pm}$ and $\bar{p}/\pi^\pm$. Other ratios such as the ratio of strangeness baryons to pions are calculated from the parameters thus obtained without any adjustment and the agreement  with  data is surprisingly good. Including the calculated ratios, the $\chi^2 /N$ is 0.9. As a byproduct  the (pseudo-)rapidity distribution of the total charged hadrons is obtained only by adjusting the parameter, $\eta_{max}$.

After a brief explanation of the blast-wave model with two freeze-outs, whose equations are used in this calculation,
a new method of doing both the chemical and thermal analysis is proposed. Result of calculation is presented which is followed by Summary.

\section{Blast-wave model with Two Freeze-outs}

It is assumed that during heavy-ion collisions a hot and dense quark-gluon plasma is formed and expands hydrodynamically. As the temperature decreases below the critical temperature hadronization occurs and the hadronic matter further expands.  At $T_{ch}$, chemical freeze-out occurs and the chemical composition of hadrons does not change below $T_{ch}$. The system further expands radially and longitudinally and cools down maintaining thermal equilibrium until the thermal freeze-out at $T_{th}$. In between $T_{ch}$ and $T_{th}$, it is assumed that all the numbers of hadrons are kept fixed, except for those involved in a process with resonances with very large cross-section such as  K$^*$ mesons or $\Delta$ baryons. 
At thermal freeze-out at $T_{th}$  the system expands longitudinally with surface rapidity $\eta_{max}$ and transversally with surface rapidity $\rho_0$, whose values determine the transverse and rapidity spectra of various hadrons.



Below a blast-wave model recently developed\cite{choi} in order to implement the two different freeze-outs is summarized because the equations in it will be used. Basically it is same as the model by Dobler {\it et al.}\cite{dobler}
except the treatment of chemical potentials. 
Cylindrical geometry of the fireball is assumed in this model and  the equation for  transverse mass spectra can be written as

In the blast-wave model with cylindrical geometry\cite{dobler}, the differential cross section of hadrons is given  from the Cooper-Frye formula as: 

\begin{widetext}
\begin{equation}
\frac{\mathrm{d} N_i^{th}}{m_T \mathrm{d}m_T \mathrm{d} y} =\frac{d_i V_{eff}}{(2\pi)^2} \int^{\eta_{max}} _{-\eta_{max}}  \mathrm{d}\eta \int^{R} _0 \mathrm{r}dr m_T \cosh(y-\eta) 
 \exp{(-\frac{m_T \cosh(y-\eta)\cosh{\rho} -\mu_i}{T})}\, I_0 (\frac{p_T \sinh{\rho}}{T})
 \label{spec}
\end{equation}
\end{widetext}
Here $\eta_{max}$  is the logitudinal rapidity at the surface and the logitudinal rapidity at $z$ is assumed to be linearly scaled from the center to the surface. $rho(r)$ is the transverse rapidity at the transverse radius $r$ is assumed to follow the linear scaling:  $\rho(r)=\rho_0 (r/R)$ where $R$ is the radius to the surface. By changing the integration variable from $r$ to $r/R$, the radius $R$ becomes a multiplication factor and can be absorbed into $V_{eff}$. 

Together with the thermal hadrons, those from the decay of resonances with higher mass should be added. The program to calculate $\mathrm{d}N_i^{res} /m_T \mathrm{d}m_T \mathrm{d}y$ including 2 and 3 body decays has been written by Sollfrank\cite{sollfrank} and used in our calculation. To get the number of hadrons from the resonance contribution, integration over $y$ is needed. It should be emphasized that integrating only over the measured range of $y$, $-y_m < y <y_m$ should be done to get the measured number of hadrons including those from the resonance decay, or extrapolation to all the rapidity range is needed. When the system is invariant under the Lorentz boost, $dN/dy \propto f(T)$ in the midrapidity region and the ratio of hadrons can be expressed as the exponetial of the difference of chemical potentials, as is done in usual statistical models.

Adding the contribution from resonance decay, one gets an equation for the total number of hadrons.
\begin{equation}
N_i =\int _{-y_m}^{y_m} \mathrm{d}y \int m_T \mathrm{d}m_T  \frac{\mathrm{d}^2 (N_i^{th}+ N_i^{res})}{m_T \mathrm{d}m_T \mathrm{d}y}
\label{number}
\end{equation}
Since the number of hadrons is Lorentz invariant, the result should be the same as using  the thermal distribution function without any Lorentz boost is used, which is essentially a statistical model. 

When the system reaches chemical freeze-out at $T_{ch}$, numbers of hadrons are parametrized by the two parameters, namely the baryon and the strangeness chemical potentials, $\mu_B$ and $\mu_S$ so that chemical potential for $i$-th particle $\mu_i$ is the algebraic sum of the two, 
\begin{equation}
\mu_i = (n_q -n_{\bar{q}} ) \mu_B /3 +(n_s -n_{\bar{s}} ) \mu_s .
\label{ch}
\end{equation}
 Below $T_{ch}$, number of each hadron species is kept fixed by neglecting the small change in hadron numbers due to  reactions with large cross sections such as $\rho \rightleftharpoons \pi +\pi $, which is called as complete chemical freeze-out(CF) compared to the case when it is included as in the partial chemical equilibrium(PCE)\cite{hirano,teaney}. Here CF is assumed for convenience of calculation. When the system reaches thermal freeze-out at $T_{th}$, chemical potential for hadron species, $i$, should be determined from the number of hadrons, $N_i$ , which is already known  at $T_{ch}$. Thus
\begin{equation}
\mu_i = T(T_{th}) \ln[ N_i  (T_{ch})
\int\int m_T \mathrm{d}m_T \mathrm{d}y (\frac{\mathrm{d}^2N_i ^{'}(T_{th}) }{m_T \mathrm{d}m_T \mathrm{d}y}) ]
\label{chempot}
\end{equation}
where the $'$ denotes that $\exp{(\mu_i /T)}$ is absent in this equation compared to Eq.~(\ref{spec}).
It should be worth to mention that the strangeness chemical potential $\mu_s$, which was one of the fit parameter at $T_{ch}$ does not appear in Eq.~(\ref{chempot}) and only the $\mu_i$ is calculated using Eq.~(\ref{chempot}).

Given the above equations, now one can describe the procedures of new method of both the chemical and thermal analysis and the method has been applied to the hadron data by ALICE collaboration\cite{alice1,alice2} in Pb+Pb collisions at   $\sqrt{s_{NN}} = 2.76 $ TeV in the rapidity windows $-0.5<y<0.5$. 

\section{A New Method of Chemical and Thermal Analysis and Results}

As was discussed earlier, in analysing the transverse momentum spectra, resonance decay is important and  the numbers of resonances  to decay are assumed to be chemically equilibrated at $T_{th}$ and calculated at $T_{th}$. In other words, the numbers of hadrons are obtained from the transverse momentum spectra by extrapolation them to zero $p_T$ using a blast-wave model with resonance contribution, which is large at small $p_T$ region. And the numbers of resonances are obtained by assuming the chemical equilibirum at $T_{th}$, which is not true. Most of the numbers of resonances should be determined at $T_{ch}$, which is unknown unless one performs the chemical analysis with the numbers calculated with a wrong assumption. Thus in order to circumbent the difficulty, iteration methos is proposed below and shown to be very successful.

The iteration  procedures of both the chemical and thermal analysis are as follows: 

\noindent (1) Assume any reasonable values for chemical freeze-out parameters such as $T_{ch}$, $\mu_B$, $\mu_S$ and the overall constant and calculate all the particle numbers including that of resonances and store them.

\noindent (2) Using the stored number for each hadron species, calculate the chemical potential from Eq.~(\ref{chempot}) and fit the transverse momentum spectrum with Eq.~(\ref{spec}) to find $T_{th}$, $\rho_0$ and the overall constant. 
Transverse momentum spectra of $\pi^+$, $\pi^-$, K$^+$, K$^-$, p and $\bar{\mbox{p}}$ measured by ALICE collaboration\cite{alice1} in Pb+Pb collisions at   $\sqrt{s_{NN}} = 2.76$ TeV in the rapidity windows $-0.5<y<0.5$ has been analysed using the blast-wave model described in the previous section. 

The transverse momentum spectra are insensitive to a value of the longitudinal rapidity at the surface, $\eta_{max}$, while the width of the $y$ or $\eta$ distribution critically depends on $\eta_{max}$. The value  $\eta_{max}$ is determined  to fit the width of the pseudo-rapidity distribution of the total charged hadrons measured by ALICE collaboration\cite{alice2} and kept fixed throughout the calculation. 

\noindent (3) From the fitted parameters at thermal freeze-out, namely $T_{th}$, $\rho_0$, $\eta_{max}$ and an overall constant together with $\mu_i$'s, one can calculate the number of the total or only the thermal hadrons either in the small rapidity window measured or in the whole range of rapidity. Please remember that the numbers of hadrons are obtained with the number of resonances calculated at $T_{ch}$, but not at $T_{th}$. In the scenario of two freeze-outs, there is no way to know the numbers of resonances to decay at $T_{th}$ and thus one has to iterate the chemical and thermal analysis to minimize the error in both analysis.

\noindent (4) Now  the number of thermal hadrons in the small rapidity interval is obtained by subtracting the resonance decayed ones from the measured value. 
Chemical analysis can be done only for the thermal hadrons, namely for pions, kaons, protons and anti-protons.  By integrating  Eq.~(\ref{spec})  over $m_{T,0} < m_T$, one gets equation for thermal hadrons to fit and it is important to use  Eq.~(\ref{ch}) for the chemical analysis. 
In this way  one gets parameters at chemical freeze-out, $T_{ch}$, $\mu_B$ and $\mu_S$ and overall constant, if needed. If one wants to fit ratios, the corresponding equation can be obtained easily from Eq.~(\ref{spec}). 

With a new set of fitted parameters at $T_{ch}$, the numbers of all the particle species have to be stored in computer.

\noindent (5) The steps from (1) to (4) are repeated until the parameters converge with smallest least squred error and  the ratios, transverse momentum spectra, and the rapidity distribution of charged hadrons are fitted simultanesouly  and consistently in one single blast-wave model.

The hadron data by ALICE collaboration in Pb+Pb at   $\sqrt{s_{NN}} = 2.76$ TeV have been analyzed following the steps described above and the resulting parameters are tabulated in Table I. The chemical freeze-out temperature is 155.7 MeV, consistent with other calculations. The ratios are shown in Fig.~1 together with data. Only the 4 ratios from the left, whose transverse momentum spectra are available, are the fitted ones.  The ratios involving strange baryons, such as $\Lambda$, $\Xi$ and $\Omega$ to pions are just calculations using the parameters obtained. 
In order to show how good is the agreement, $\chi^2 /N$ is calculated including all the ratios in the figure, which is 0.9.
Here there is no discrepancy for protons and the strangeness fugacity is kept fixed to be 1.
The fitted transverse mass spectra of measured hadrons are drawn in Fig.~2, together with ALICE data up to $m_T =3$ GeV. Except for small discrepancy in large $m_T$ region of protons and anti-protons, 
the agreement is very good. which is believed to come from the equal treatment of the spectra with different magnitudes of pions and protons. Use of weighted fitting procedure to make the magnitudes of different hadrons eqaul will remove this discrepancy at large $p_T$ of massive hadrons.

In Fig.~3, the pseudo-rapidity distribution calculated is plotted together with the data by ALICE collaboration\cite{alice2}. As was mentioned above, the value for $\eta_{max}$ is varied to fit the width of the measured $dN_{ch}/d\eta$, and the resulting $dN_{ch}/d\eta$ from the parameters to fit the transverse momentum spectra agrees with data very nicely. This shows that the assumption of uniform temperature inside the fireball is reasonable.

\begin{table}
\caption{Chemical(C.F.) and Thermal freeze-out(T.F.)  parameters fitted to data by ALICE Collaboration. T and $\mu$'s are in MeV.}
\begin{tabular} {|c|c|c|c|c|c|c|clcl}\hline
    & T & $\mu_B$   &  $\mu_S$ & $\rho_0$ & $\eta_{max}$&$\chi^2/N$ \\ \hline

C.F.  & 155.7 &  1.7  &  0.14 & 1.3 & 6.8 & 0.9  \\ \hline

T. F.    & 112.0 &   &   &  1.3 &  6.8 & 3.8 \\ \hline

\end{tabular}
\end{table}

\begin{figure}

\begin{center}
\resizebox{90mm}{!}{\includegraphics[angle=-90]{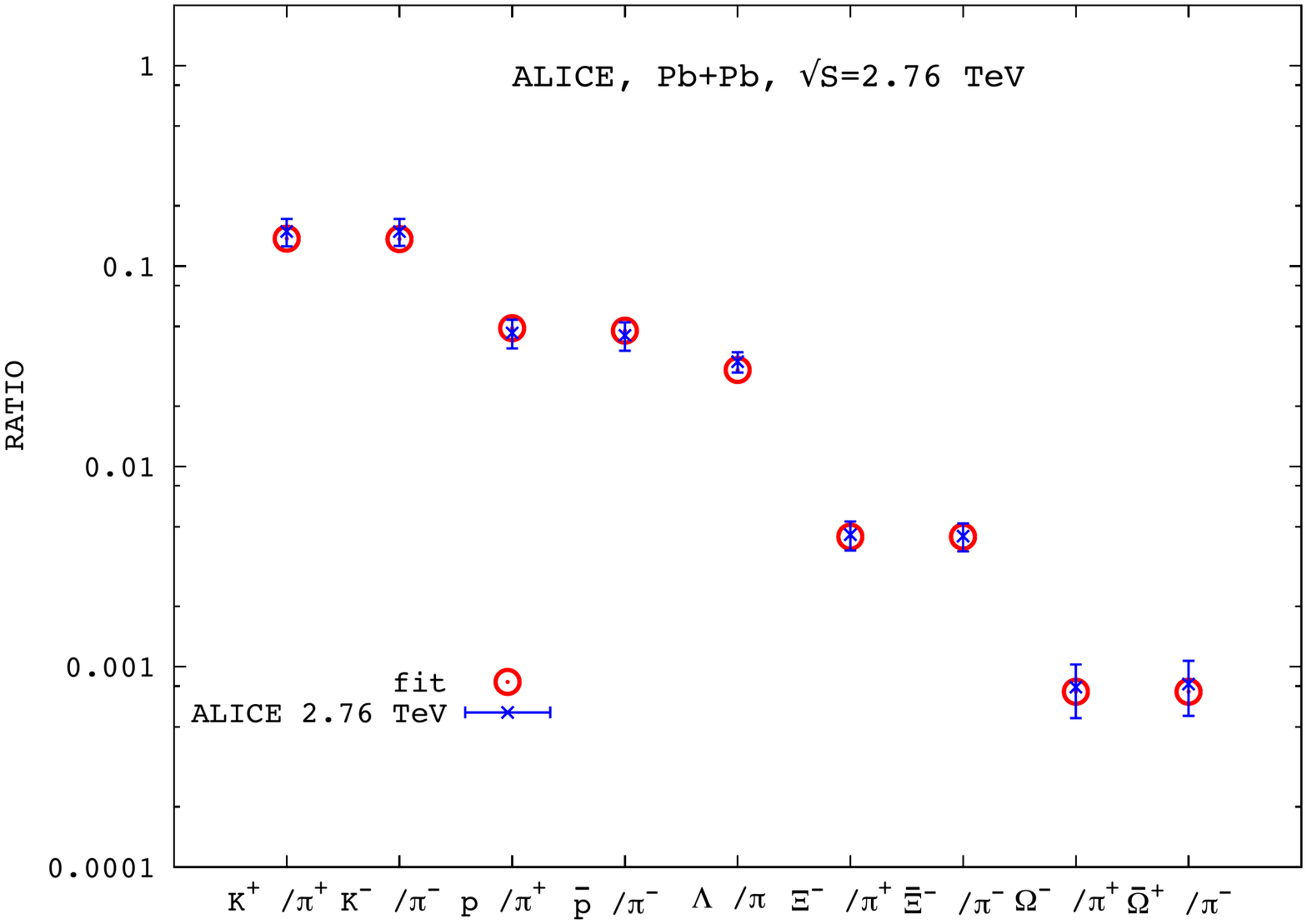}}
\end{center}
\caption{Ratios among hadrons by ALICE collaboration\cite{alice1} in Pb+Pb collisions at   $\sqrt{s_{NN}} = 2.76 $ TeV and result of fitting. It should be emphasized that only ratios involving pions, kaons, protons and anti-protons are fitted. Other ratios are calculation from the resulting parameters.}
\label{aliceratio}
\end{figure}

\begin{figure}

\begin{center}
\resizebox{90mm}{!}{\includegraphics[angle=-90]{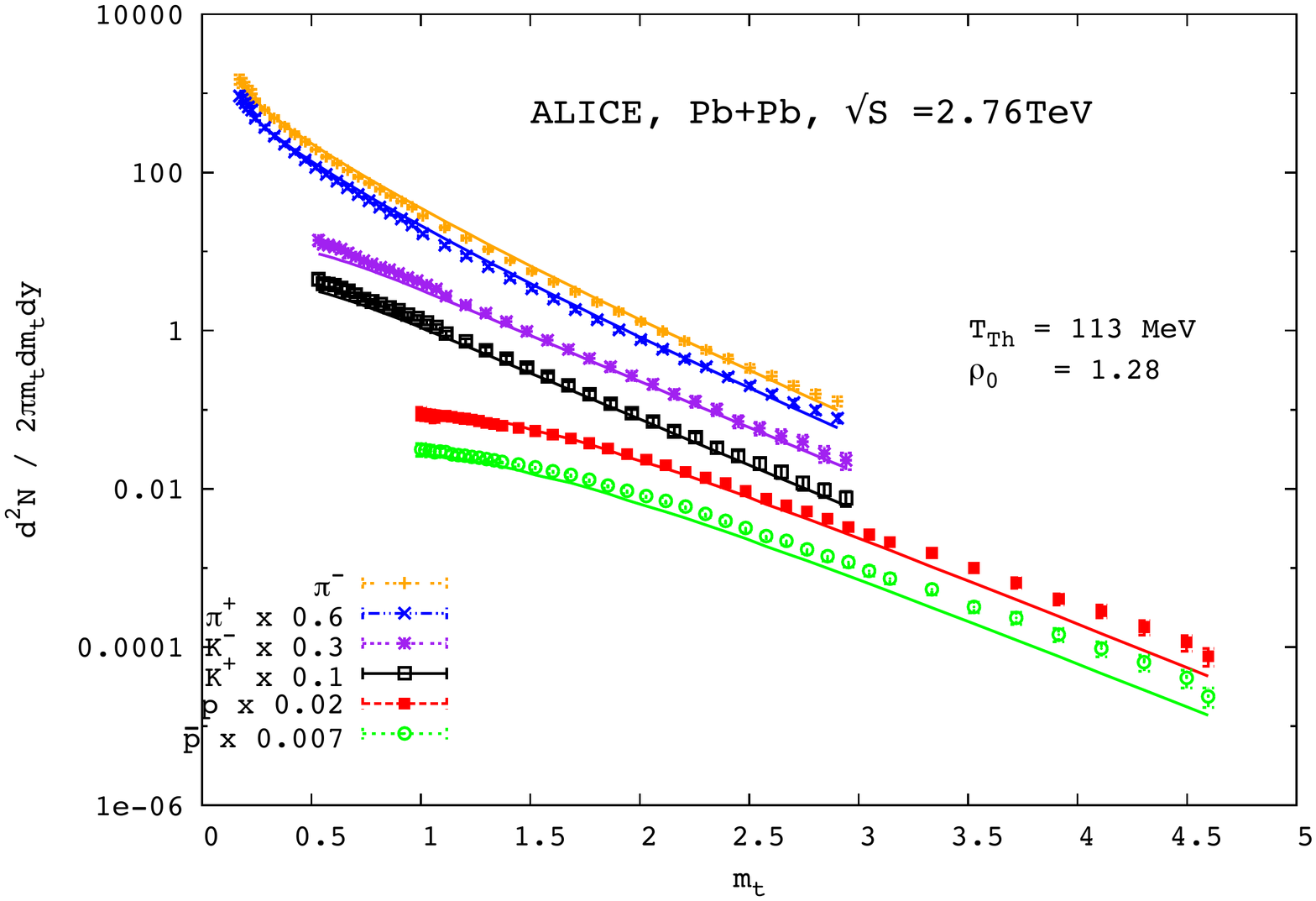}}
\end{center}
\caption{Transverse mass spectra of measured hadrons by ALICE collaboration\cite{alice1} in Pb+Pb collisions at   $\sqrt{s_{NN}} = 2.76$ TeV and fitted curves.}
\label{alicemt}
\end{figure}

\begin{figure}

\begin{center}
\resizebox{90mm}{!}{\includegraphics[]{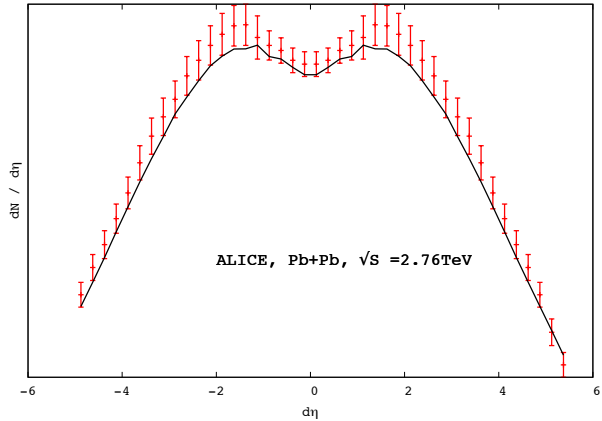}}
\end{center}
\caption{Psuedorapidity distribution of charged hadrons by ALICE collaboration\cite{alice2} in Pb+Pb collisions at   $\sqrt{s_{NN}} = 2.76 $ TeV. Only $\eta_{max}$ has been adjusted to fit the width.}
\label{dNdeta}
\end{figure}

\section{ Summary }

A new iterative method of analyzing both the ratios and the transverse momentum spectra of measured hadrons within a blast-wave model with two freeze-outs is presented and applied to  ALICE data to show surprisingly good agreement. Even though only those hadrons whose transverse momentum spectrum is available are included in the analysis, the $\chi^2 /N$ is rather small even when the calculated ratios involving other strange baryons are included in $\chi^2 /N$. The ratios involving protons and anti-protons are nicely fitted. Even the pseudo-rapidity distribution of the total charged hadrons is nicely reproduced only by adjusting the value for $\eta_{max}$ from the parameters fitted from the transverse momentum spectra, which shows that the assumption of the uniform temperature inside the fireball is reasonable. 

The chemical freeze-out temperature is same as obtained in the literatures. This method may make considerable difference from the usual methods when $dN/dy$ is not invariant under the Lorentz boost and thus the calculation using the same method for RHIC data is under way.

 \acknowledgments

{K.S. Lee is supported by Basic Science Research Program through the National Research Foundation of Korea(NRF) funded by the Ministry of Education, Science and Technology(2011-0010433). Suk Choi is supported by the Rare Isotope Science Project of Institute for Basic Science funded by the Ministry of Science, ICT and Future Planning (MSIP) and the National Research Foundation (NRF) of the Republic of Korea under Contract 2013M7A1A1075764. K.S. Lee expresses thanks to T. Hirano and U. W. Heinz for helpful discussions. }



\begin{references}

\bibitem{braun1} P. Braun-Munzinger, K. Redlich, J. Stachel, in Quark Gluon Plasma 3, ed. by R.C. Hwa and X.N. Wang (World Scientific Pub., 2004); and references therein.
\bibitem{andro} J. Stachel, A.  Andronic, P. Braun-Muzinger and Redlich, 
\bibitem{becat2} F. Becattini, Grassi,  arXiv:1405.0710
\bibitem{manninen} J. Manninen and F. Becattini, Phys. Rev. C{\bf 78}:054901 (2008)
\bibitem{rafelski} J. Letessier and J. Rafelski, Eur. Phys. J. A {\bf 35}, 221 (2008)
\bibitem{pbm} A. Andronic, P. Braun-Muzinger and J. Stachel, Phys. Lett. B{\bf 673},142 (2009).
\bibitem{becat} F. Becattini, P. Castorina, A. Milov and H. Satz, Eur. Phys. J. C{\bf 66}. 377 (2010).
\bibitem{cleymans} J. Cleymans, {\it et. al.  }, Phys.Rev. C{\bf 71}, 054901 (2005).
\bibitem{siemens} P.J. Siemens and J.O. Rasmussen, Phys. Rev. Lett. {\bf 42}, 880 (1979).
\bibitem{lee} Kang S. Lee, U. Heinz, and E. Schnedermann, Z.
Phys. {\bf C 48}, 525 (1990).
\bibitem{sollfrank} E. Schnedermann, J. Sollfrank, and U. Heinz, Phys. Rev. {\bf
C 48}, 2462 (1993).
\bibitem{dobler} H. Dobler, J. Sollfrank, and U. Heinz, Phys. Lett. B{\bf457}, 353 (1999).
\bibitem{song} H. Song, S.A. Bass and U. Heinz, Phys. Rev. C{\bf 83}, 024912(2011).
\bibitem{jeon} Ryu {\it et. al.}, nuch-th/1210.4588, Proceedins of the QM2012.
\bibitem{hirano1} S. Takeuchi {\it et al.}, Phys. Rev. C{\bf 92}, 044907 (2015).
\bibitem{choi} Suk Choi and Kang Seog Lee, Phys. Rev. C {\bf 84}, 064905 (2011).
\bibitem{alice1} B. Abelev {\it et al.}, ALICE Collaboration, Phys. Rev. C{\bf 88}, 044910 (2013).
\bibitem{alice2}  ALICE Collaboration, Phys. Lett. B{\bf 726}, 610 (2013).
\bibitem{hirano} T. Hirano and K. Tsuda, Phys. Rev. C{\bf 66}, 054905 (2002).
\bibitem{teaney} D. Teaney, nucl-th/0204023.

\end{references}
\end{document}